# Analysis of Aperture Evolution in a Rock Joint Using a Complex Network Approach[1]

Hamed .O.Ghaffari & Mostafa.Sharifzadeh

*Department of Mining & Metallurgical Engineering, Amirkabir University of Technology, Tehran, Iran*

**Abstract:** In this study, we develop a complex network approach on a rough fracture, where evolution of elementary aperture during translational shear is characterized. In this manner, based on the Euclidean measure, we make evolutionary networks in two directions (in parallel and perpendicular to the shear direction) and on the measured apertures' profiles. Evaluation of the emerged networks shows the connectivity degree (distribution) of networks, after a transition step; fall in to the stable states which are coincided with the Gaussian distribution. Based on this event, we present a model in which evolving (decaying) of networks are accomplished using a preferential detachment (based on certain probability) of edges. Also, evolving of cluster coefficients and number of edges displays similar patterns as well as are appeared in shear stress and dilation changes, respectively.

## 1. Introduction

Understanding of rock joint behaviors, either in single or swarm form, under the several natural or artificial forces has allocated numerous researches during the evolution of rock mechanics field. Scientists have shed light on the several approaches to capture interacted rock joint(s) manners as well as analytical solutions to hybrid numerical methods and the methods based upon indirect investigations such as fractal , statistical or natural based computing methods [1]-[3]. Rock joint performance as a result of collective behavior of the constructed elements (say fraction/pixel in each surface), interacting with each other, determines nonlinear picture of a changeable system. Obviously, one cannot predict the rich behavior of the whole by merely extrapolating from the treatment of its units [23].

---

This is a prevalent gesture of complex systems. The success in describing of interwoven systems using physical tools as a major reductionism is associated with the simplifications of the interactions between the elements so that complexity reduction is a rescue pathway to regulation of approximated analyses of collective particles having swing states, complicated structures, and diversity of relations among elements. Dissection of phenomena under investigation into a list of interacting unites- which are building blocks (granules) of system- connected by pair-wise connections, can be revealed in complex networks, which provide a mathematical framework to analysis of wide range of complex systems. Picturing, modeling and evaluation in a simple and intuitive way are some of the discriminated features of complex networks [4],[5].Complex networks have been developed in the several fields of science and engineering for example social, information, technological, biological and earthquake networks are the main distinguished networks [6]-[9].

On the other hand, to catch on Hydro-mechanical and mechanical behavior of a rock joint, domination on to the surface morphology evolution as well as aperture is irrefutable. In this way, characterization of aperture of a rock fracture has been realized by using several techniques such assigning of probability distributions (random field theory) [1], [2] and – [11], [12] correlation length [14], fractals [13] and spatial correlation (semi-variorums) [3]. In this study, we provide a complex network approach on the aperture evolution during transitional shear accounting in two separated directions of shear while the disclosed networks involve the inherent difficulties such dynamical concepts of components, changing the wiring diagrams and meta- complications. Also, upon the consequences we present an algorithm, offering a general view of transitions of an elementary aperture distribution, which its main core is associated by preferential detaching of similar granules. In addition to these procedures, the mechanical properties and hydraulic conductivity of the being joint are compared with the networks properties.

The first part of this article covers the employed method, some preliminary aspects of networks measurements, and the proposed algorithm on the real results accompanying an analytical solution for the algorithm. The next section includes the results obtained from the experimental tests on a rock joint and the arranged complex networks. In discussion part, we investigate other types of similarity measures -to construction of networks edges-which exhibit other sides of a rock joint evolution, the affections of contacts zones and their variations at the successive displacements.

**2. Method**

A network (graph) consists of nodes and edges connecting them [20]. In this study our aim is to setting up of a network approach on the measured apertures' profiles so that characterization of the appropriate aperture behavior under the successive shear displacements is accomplished. To set up a network on the aperture patterns, we consider each aperture profile[2] as a node, so named profile-profile network (Fig.1). To make edge between two nodes, a relation should be defined. In this study we focus on the similarity measure between nodes while two mathematical measures are employed: Euclidean distance as a core of our networks and Chebyshev (the $L_\infty$ metric) as a complementary metric, which are given as below, respectively [24]:

$$d_{Euc.} = \sqrt{\sum_{p,q=1:n_p} (p(x_1, x_2, ..., x_n) - q(x_1, x_2, ..., x_n))^2} \quad , \tag{1}$$

$$d_{Che.} = \max_i(|x_i^p - x_i^q|) = \lim_{k \to \infty} (\sum_{i=1}^{n} |x_i^p - x_i^q|^k)^{\frac{1}{k}} \quad , \tag{2}$$

where $p$ and $q$ are the $i^{th}$ profiles and $x_k^i (i \in \{p \vee q\})$ shows $k^{th}$ element from the $i^{th}$ profile. It must be noticed the Chebyshev distance returns the maximum distance between elements in the assumed profiles. When $d \leq \xi$ an edge among two nodes is created. As it can be seen the emerged networks upon the mentioned way are undirected networks. The threshold $\xi$ depicts error level, usually can be assumed as 5-20 percent of maximum $d$ (Here we put =5 for Euclidean distance). In the other view and based on granularity [17], [18] of the collected information, choosing of such constant value may be associated with the current accuracy at data accumulation where after a maximum threshold the system (here apertures evolution) lose its dominant order. In the parallel with this discussion, scaling or categorization of the appeared apertures (points) shows a similar procedure so that distribution transition from a binomial state (when we don't consider variation of apertures values) to other multi-bins cases (which are similar to the feature) can be observed. This proves how discritization of an event displays different and sometimes contradictive faces of dynamical systems.

---

[2]-X-profiles: apertures' profiles parallel with the Y-axis (perpendicular to the shear direct) and Y-profiles are parallel with the shear direct.

Let us introduce some properties of the undirected networks: clustering coefficient ($C$) and the degree distribution ($P(k)$). The clustering coefficient describes the degree to which $k$ neighbors of a particular node are connected to each other. Our mean about neighbors is the connected nodes to the particular node. To better understanding of this concept the question "are my friends also friends of each other?" can be used. In fact clustering coefficient shows the collaboration (or synchronization and tendency) between the connected nodes to one. Assume the $i^{th}$ node to have $k_i$ neighboring nodes. There can exist at most $k_i(k_i-1)/2$ edges between the neighbors (local complete graph). Define $c_i$ as the ratio

$$c_i = \frac{actual\ number\ of\ edges\ between\ the\ neighbors\ of\ the\ i^{th}\ node}{k_i(k_i-1)/2} \quad (3)$$

Then, the clustering coefficient is given by the average of $c_i$ over all the nodes in the network:

$$C = \frac{1}{N}\sum_{i=1}^{N} c_i. \qquad (4)$$

For $k_i \leq 1$ we define $C \equiv 0$. The closer $C$ is to one the larger is the interconnectedness of the network. The connectivity distribution (or degree distribution), $P(k)$ is the probability of finding nodes with $k$ edges in a network. In large networks, there will always be some fluctuations in the degree distribution. The large fluctuations from the average value ($<k>$) refers to the highly heterogeneous networks while homogeneous networks display low fluctuations. The mentioned properties of large networks (and so other measures [4], [5] and [7]) make the relatively analysis of the possible statistical properties of such networks. By using these attributes, we can manifest the underlying laws that govern the evolution of the complex networked systems.

There are different types of the networks models which have been developed based on specific events in the real world, for instance the Erdos-Renyi (random)[20], the small-world ( Watts-Strogatz model[8],[10]) , and the scale-free (Albert-Barabasi model) models [5],[6]. In this part of this study and upon the observed emerged behavior of the covered networks on the apertures' profiles (section results), we present a simple model which considers the edges detachments (decaying of initial similarity patterns) of the networks. So, using a continuous analysis the behavior of the proposed algorithm is proved. For simplicity, our model doesn't take in to account the nodes decaying (as it comes out in the X-profiles evolving) and attempt to capture the nearly possible mechanism(s) that govern the evolution of network topology, guided by the real information involved in the  degree distribution. Our algorithm is another changed version of the limiting case of the scale-free model [5], [6] where the decaying of network is ignored. The steps of the algorithm are as follows:

1) Starting with *N* nodes (number of profiles) and like real initial state of a joints 'profiles, construct a fully connected graph (lattice network) where all of nodes are connected except a few of them.
2) At each time step, select a node uniformly and detach the edge which its end point (node *i* with the degree $k_i$) is selected with a preferential probability ,is given as:

$$\Pi(k_i) = \frac{k_i}{\sum k_j},$$

where the sum in the denominator goes over all nodes in the system except the source node. The change rate of node connectivity has two components: the first describes the probability that node $i$ is chosen randomly as: $\Pi_{random}(k_i) = 1/N$ and the second is related with $\Pi(k_i) = k_i / \sum k_j$, regulating the probability that an edge beginning from a randomly selected node is detached from node $i$ :

$$\frac{\partial k_i}{\partial t} = -(A \frac{k_i}{\sum_{j=1}^{N} k_j} + \frac{1}{N}) \qquad (5)$$

which gives the time dependence of the degree $k_i$ of an assumed node $i$ while under this assumption that $k_i$ is a continuous real variable . Considering $\sum k_j = N^2 - N - 2t$ and the variation of connectivity during time step is $\Delta(k) = -2$, we obtain $A \approx (2N - 1)$, so since $N \gg 1$ then $A \approx 2N$. Replacing them at Eq.(5) and solving of this equation is led to the approximated form :

$$k_i(t) \approx (N - \frac{2t}{N})(1 - \frac{2t}{N^2})^N \qquad (6)$$

As one can see the emerged equation shows the similar behavior as it is appeared in real evolution of X or Y apertures' profiles. The main difference with the real observations (results section) is in the very soft transition while occurs at the delayed time. To capture this state, we can change the number of random selection of nodes to the $m$ times at each step. Fig.2a and b show the evolution of edges frequency obtained from the mentioned algorithm ($N$=90) over the different time step. Fig.2c and d illustrate the variations of the edges ($\sum k_i$) and $k_i(t)$-using Eq. (6)-during 2500 and 8000 time steps, respectively. In the other process, by several values of N, Eq. (6) has been portrayed (Fig.3). As it can be seen after $t \approx N^2$ time steps the system reaches a state which all nodes are isolated. It should be noticed that Eq. (6) is valid only for $t > t_{int.}$ where $t_{int.}$ is the time when node $i$ was selected for the first time as the origin of an edge. So, all nodes follow the aforesaid dynamics after $t \geq N$.

Also, as if our system cannot capture all of the appeared states of the real complex networks, but may give an overall view on the designated networks on the apertures alterations: "*destroying of the high similar profiles (at least in initial steps) associated with a preferential probability*". It can be proved that removing of the most highly connected nodes at each step bears the most damaging to the integrity of the system but in this case the system doesn't show like behavior in the edges distribution and linear rate of nodes removal [5]. Considering this point that in Y-profiles against X-profiles the number of active (non-isolated) nodes shows a constant state, the dominant mechanism can be conjectured as a combining of two processes on the overall profiles (real joint) . One may change the constructive parts of Eq. (5) to reach more exact results; for instance, considering non-linear core in the preferential probability or selection/detachment depends on time passing.

## 3. Results

In this part, we focus on the experimental results and mapping them in to the complex networks. Regard this point that the network anatomy is so important to characterization (because structure always affects function) our aim is underlined to find out the possible relations between the constructed networks properties and the current mechanical / hydro-mechanical properties of a rock joint which is under a constant normal stress and the successive shear displacements.

The rock material was granite with the weight of 25.9 and uniaxial compressive strength of 172 Mpa. An artificial rock joint was made at mid height of the specimen by splitting using special joint creating apparatus, which has two horizontal jacks and a vertical jack [1],[15]. The sides of the joint are cut down after creating joint and its final size is 180 mm in length, 100 mm in width and 80 mm in height. Using special mechanical units the different mechanical parameters of this sample were measured. A virtual mesh having a square element size of 0.2 mm spread on each surface and each position height was measured by the laser scanner. Also using a special hydraulic testing unit is employed to allow linear flow experiments (parallel with shear direction) to be governed while the rock joint is undergoing normal or shear loading. The details of the procedure can be followed in [15], [16]. The hydraulic conductivity and hydraulic aperture are given by Darcy's law:

$$Q = K_h . A . i , \qquad (7)$$

and assuming the joint surfaces are as two smooth parallel plates ,the flow rate and hydraulic conductivity can be written as below:

$$Q = \frac{g \, e_h^2}{12.\upsilon} . (w \, e_h) . i , \qquad (8)$$

$$K_h = \frac{g \, e_h^2}{12.\upsilon} , \qquad (9)$$

where $Q$, $A$, $i$, $K_h$, $g$, $e_h$, $\upsilon$ and $w$ are the volumetric flow, area, hydraulic gradient, hydraulic conductivity, the gravity acceleration($m/s^2$), hydraulic aperture, kinematic viscosity of fluid and the width of the specimen ,respectively.

In this study, we consider only the evolving of apertures under constant normal stress and regular translational shear in which the lower surface has fixed position and upper one is displaced (Fig.4). By employing a threshold value-$d \leq 5$- in Euclidian distance (Eq. (1)) and setting up a pre-designated complex network (Fig.1a) on the X-profiles, gradual changes of the adjutant matrix form of the appeared networks can be inferred (Fig.5). Fig.5 demonstrates after a phase transition step the similarities' patterns are constrained to the adjacency of each profile. The neighborhood radius –in the final stages of disruption- changes from 2 to 20 pixels (0.4-4 mm), except for boundary profiles. The concentration of similarities takes place around 5-10 pixels where lower and higher values sort in a symmetric shape which disclose a Gaussian distribution.

Portraying total number of edges, number of active nodes (non-isolated sites) and clustering coefficient, during transitional shear stress, reveals others properties of the assigned networks. At X-profiles, decreasing of the edges after SD=1 (Fig.6 a) is coincided with a transition point (interval) which shows the system after an abruptly falling, goes towards a stable state. This behavior, called phase transition, is a usual and current behavior in several natural systems such as physical, social and economical systems [18], [21] and [22].

In fact, elicitation , prediction and finding out of effective parameters of phase transition step(s) are one of the main stages in this track where system goes from a stable (or unstable) to another stable (or unstable such as social revolutions) and order (disorder) parameter(s) characterizes these transformation occasions. If we transform $K$ to the reverse case and re-plot the overall changes in a log-log coordinates, the general form of a continuous phase transition is appeared. It must be noticed that the phase alteration in this case is not an order to disorder (or reveres case) but is a soft transition from an order to the semi-order case where after 4 mm the system reaches a semi-stable state. We can designate a sigmoid function [22] and with some mathematical manipulation, is followed as below:

$$k = 2\times 10^5 \times (1+e^{-\beta(1+\frac{LnSD}{1+\delta LnSD})}) \qquad (10)$$

where $\beta$ and $\delta$ are the regulator parameters which determine the declining rate (Fig.7). It should be reminded the proposed algorithm (section 2) represents a similar graph for edges descending.

A similar behavior can be followed in the appropriate Y-profiles where the number of active nodes has a constant value and doesn't show an intermediate step in declining procedure (discontinuous transition). Here, the order parameter is selected as the number of edges which in general form is concurred with the contact percent or normal displacement versus shear displacements (Fig.6.d). This implies that the decreasing of contact zones induces lower similarities between profiles (unmated) either in X or Y directions. However, by indenting of contact zones and diminishing of percolation clusters, the proper networks with the stair-profiles expose interesting property where in first look have a contradictory manner: emerging of growing networks (refer to the discussion section).

Gradual changes of connectivity distributions, either in X or Y profiles, reveals a similar transitional behavior: transformation from a nearly single value function to a semi-stable Gaussian distribution (Fig.8). In parallel profiles of networks with shear direction the neighborhood radius shows a lower interval rather than X-profiles, so-in agree with connectivity numbers changes- transition to a semi-stable stage is occurred with more convergence rate. In other word, similarities of Y-profiles occur in higher constraints and readily (faster than opposite option). The possible reason can be inferred from the lower resistance against transitional shear (and also easy leading of fluid flow), results more loosing of possible contacts –increasing percolation clusters-and abrasion of asperities in the Y-profiles.

The complementary results using calculation of clustering coefficients emphasis that the inverse values of $C$ during process, in X-profiles, hand out like pattern as ones came out in the variations of shear stresses (Fig.9 a and c) while in the same manner and on Y-profiles resembles with the increment of hydraulic conductivity values (Fig.9 b and d). The coinciding of $1/C_Y$ with the changes of the hydraulic conductivity rather than the dilation behavior proves that the joint un-matching (or increment of mean aperture) is not only singular parameter in the fluid flow, as if reduction of joint roughness and the entire ensembles of the similarity behaviors of asperities are other possible agents.

Also, these results indicate that lower synchronization in X-profiles causes great resistance against shear stress and higher orchestration at Y-profiles bears minimum hydraulic conductivity, particularly at a step that interlocking of (SD=1) asperities and simultaneously decreasing of flow pathways take place. It should be noticed that the sensitivity of the clustering coefficients to the scaling is very high, for instance with 5 and 10 times of virtual meshes (1-2 mm), the details of the $C$ fluctuations are omitted, however, the general form of variations is preserved as it is resulted at real experimental outputs [19].

The aforesaid discussion can be perused with more details within the inter-structure of the networks. In first case, consider number of active nodes is constant i.e. $\frac{\partial N}{\partial t}=0$ (assuming shear distance has a direct relate with time) with the constraints $\frac{\partial K}{\partial t}>0$ and $\frac{\partial C}{\partial t}>0$ that indicates new edges are added to the system in the manner that the attached edges increase the collaboration of nodes, especially when $\frac{\partial C}{\partial t}>\frac{\partial K}{\partial t}$ the connections are concentrated in to the "friends" nodes (rewiring within collaborated clusters). In other word, the synchronization between the neighborhoods nodes is promoted. In another aspect, when we have $\frac{\partial C}{\partial t}\approx\frac{\partial K}{\partial t}$, implies on the uniform edges growing between vertices. For a reverse position ($\frac{\partial C}{\partial t}<\frac{\partial K}{\partial t}$), the most added edges extends in out of friends nodes ranges, i.e. the nodes are linked to new other sites. A similar argument can be followed on the other cases [19].

## 4. Discussion

In this section, we investigate the sensitivity of the profiles distance to the predefined error level, the results based on Chebyshev distance and exploring of networks due to the contact profiles. Selection of error level as a critical factor in elicitation of networks plays a prominent role. For instance the variation of edges frequency at SD=20 mm depends on threshold levels ($(0.5\leq\xi\leq 8)$ pursues an initial exponential, at elementary values, to a Gaussian at middle values and with increasing of error level looses the latter distribution (Fig.10). Nevertheless, we can follow like procedure on the connectivity distribution-time which changes from a single-scale distribution (closer to single value function) to a semi-stable Gaussian. In fact, the shape of these distributions might result from the presence of constraints limiting the number of links when connections are costly. In this sense, the exponential decays (rising) or sharp cutoffs would be a result of highly wiring [7].

As before mentioned each vertex or a cluster of nodes in a network presents one or set of properties which are connected to each other based on their similarity, functional, spatial or temporal proximity properties. These relations are the main foundation of the information packaging or elicitation of best briefed "granules" where such information granules uncover and constitute the overall and collective behavior of a complex system. Elicitation of such packages can be carried out using different, say, similarity measures which will be depends upon the problem space. In any information system several parameters can be recognized, either in statistical or dynamical systems. These attributes if have same types form a homogenous space else a heterogeneous field is produced. In our problem, the nature of attributes is apertures which create a multi-dimensional Euclidean space. The Euclidean metric can give very different results when the scale of a variable is changed .This problem may be elevated by the normalization of all attributes (a weighted Euclidean norm) .Although the Euclidean distance is the most common and popular metric in formation of granules but other similarity measures exist.

The employing of these relations is tied with the problem being solved. Here, we investigate two measurements. In first case, the Chebyshev distance (Eq. 2) which is a metric related to Euclidean and Manhantan distances is used. In Manhantan distance, the distance is computed by summing the absolute value of the difference of individual terms $(d(p,q) = \sum_{i=1}^{n} |x_i^p - x_i^q|)$ which express the minimum similarity is not always in agreement with Euclidean, especially when there are some restrictions on freely movements in the given space. The results shows similar results as one can see in the usage of Euclidean distance with some softer behavior in the initially distributions and transition interval (Fig.11) which changes from a nearly power-law ($N(k) \propto k^{-\gamma}$) to a single-scale distribution (Gaussian). The significant reason can be originated from the Chebyshev definition which the distance fall in to the maximum apertures differences, therefore, can't involve the details of the apertures details.

In other process and based upon a modified version of binary distance measure, the role of contacts zones is distinguished. Instead of pairwise comparison of elements, the Euclidean distance is utilized while percolating clusters and contact pixels are transformed in to the 0 and 1, respectively (Fig.12). Accomplishing of the profiles networks exhibits the new type of contacts profiles evolutions: Growing networks either in X or Y profiles (Fig.13).At initial steps of the evolving, neither active nodes nor edges are appeared. This is due to the high fluctuations of contact profiles which are in the stair like shapes. The variation of connectivity distribution shows a transition from an exponential to another type of one. Particularly, for X-profiles, the emanation of three distinguished zones is clear (Fig.13 a-incent). Increscent of displacement induces a lower instability of contacts profiles which impel more similarity (Fig.13 c and d). However, for instance in X-profiles, after a step (SD=6) developing of new sites are stopped while the growing of edges is continuing. In addition to this event at $\frac{\partial N}{\partial t} = 0$ (after SD=5 –compare with Fig.9a and c), we have $\frac{\partial C}{\partial t} < \frac{\partial K}{\partial t}$ proves the promoting of connectivity with other nodes (are not parts of collaborative nodes). This interval is coordinated with the reduction of roughness represents even though the percentage of contacts gets in to a nearly constant value but the similarity of contact profiles are continuously added (the coherent variation of contacts patterns). It should be ignoring of contacts is equivalent with distinguishing of percolating areas which in profile form and based on the mentioned measurement display like behavior of networks related to the contacts profiles.

## 5. Conclusions

The success in describing of interwoven systems using physical tools as a major reductionism is associated with the simplifications of the interactions between the elements where there is no possible vagueness. Employing of statistical mechanics tools gives a good framework for analysis of these systems. Also, possible relations between the building blocks of complex systems can be revealed in complex networks, which describe a wide range of systems from social systems to grid power systems, earthquake networks to other physical systems, World Wide Web to citation of papers.

From this perspective and by considering of the complicated behavior of a rough fracture which was under the shear stress, a network associated with a popular similarity measure was designated at two separated directions of the shearing. The networks properties shed light a suitable coordination with the empirically obtained mechanical and hydro- mechanical characters of the being joint. Recognition of phase transition step, reforming of inter-structures evolutions, the weight of perpendicular profiles against resistance and procuring of great order in parallel profiles are some of the benefits of the captured networks. Also, to highlight and indent the contacts functionality another bi-partite schema of profiles as dynamical nodes were taken in to account. The results emphasized on the formation of contacts clusters (in pick point at shear stress-shear displacement plot) which their similarities were increased with more ate in X-profiles than the opposite direction, representing growing networks just unlike the former case.

Based on the observations of the decaying apertures' profiles networks, we proposed an algorithm, regard to the overall performance of the real networks and transformation of connectivity distribution, which the edges break out were accomplished by preferential detachments. This may stresses on this point that ,without considering of nodes demolishing, the starting point of links cutting are more closer to the high dense nodes rather than to others. Definitely, the represented indirect modeling of the evolving networks is only a start point which can be extended with high preciseness and regarding more details.

**References**

1. Sharifzadeh M. Experimental and theoretical research on hydro-mechanical coupling properties of rock joint. Ph.D. thesis, Kyushu University, Japan; 2005.

2. Adler MP and Thovert JF. Fractures and fracture networks. Kluwer Academic; 1999.

3. Koyama T. Numerical modeling of fluid flow and particle transport in rock fractures during shear, PhD thesis ,Royal Institute of Technology (KTH),Stockholm, Sweden;2005.

4. Newman M. E. J. The structure and function of complex networks, SIAM Review 2003; 45(2): 167- 256.

5. Albert R., Barabasi A.-L. Statistical mechanics of complex networks. Review of Modern Physics 2002; 74, 47–97.

6. Barabasi A.L, Albert R, Jeong H. Mean–field theory for scale-free random networks, 1999.http://www.arxiv.org/condmat/9907068v1

7. Bomer K, Sanyal S, Vespiganani A. Network's science .In: Annual review of information science & technology 2007; 41(12):537-607.

8. Watts DJ, Strogatz SH. Collective dynamics of small-world networks. Nature 1998; 393:440-442

9. Abe S, Suzuki N. Complex network description of seismicity . Nonlin process Geophys 2006;13:145-150

10. Strogatz,S.H. Exploring complex networks, Nature. 2001; Vol .410:268-276.

11. Hakami E, Einstein H H,Genitier S , Iwano M. Characterization of fracture apertures-methods and parameters. In: Proc of the 8th Int Congr on Rock Mech, Tokyo, 1995: 751-754.

12. Lanaro F, Stephansson O.A. Unified model for characterization and mechanical behavior of rock fractures. Pure Appl Geophys,2003;160:989-998

13. Lanaro F.A. Random field model for surface roughness and aperture of rock fractures. Int J Rock Mech Min Sci, 2000, 37:1195-1210.

14. Brown SR, Kranz RL , Bonner BP. Correlation between the surfaces of natural rock joints, Geophys Res Lett, 1986; 13(13):1430-1433.

15. Sharifzadeh M, Mitani Y, Esaki T, Urakawa F. An investigation of joint aperture distribution using surface asperities measurement and GIS data processing. Asian Rock Mechanics Symposium (ARMS3), Mill Press, Kyoto, 2004:165–171.

16. Sharifzadeh M, Mitani Y, Esaki T. Rock joint surfaces measurement and analysis of aperture distribution under different normal and shear loading using GIS , Rock Mech. Rock Engng. 2006; DOI 10.1007/s00603-006-0115-6.

17. Ghaffari O.H, Sharifzadeh M, Shahrair K. and PedryczW. Application of soft granulation theory to permeability analysis, Int J Rock Mech Mining Sci, 2008;doi:10.1016/j.ijrmms.2008.09.001.

18. Ghaffari O.H, Sharifzadeh M, Pedrycz W. Phase transition in SONFIS and SORST, In RSCTC 2008, LNAI 5306, eds. C.-C. Chan et al. 2008:339 – 348.Springer-Verlag: Berlin Heidelberg.

19. Ghaffari O.H. Applications of intelligent systems and complex networks in analysis of hydro- mechanical coupling behavior of a single rock joint under normal and shear Load. M.Sc Thesis, Amirkabir University of Technology, Tehran, Iran; 2008.

20. Wilson RJ. Introduction to graph theory. Fourth Edition: Prentice Hall, Harlow; 1996.

21. Millonas M.M. Swarms, phase transitions, and collective intelligence, 1993: http://arxiv.org/abs/adap-org/9306002.

22. Levy M. Social phase transitions. Journal of Economic Behavior& Organization 2005; 57:71–87.

23. Hakan H. Synergetics: An Introduction. Springer, New York; 1983.

24. Bezdek JC, Keller J,Krisnapuram R , Nikhil RP. Fuzzy models and algorithms for pattern recognition and image processing. Springer;2005.

**Analysis of Aperture Evolution in a Rock Joint Using a Complex Network Approach-Figures**

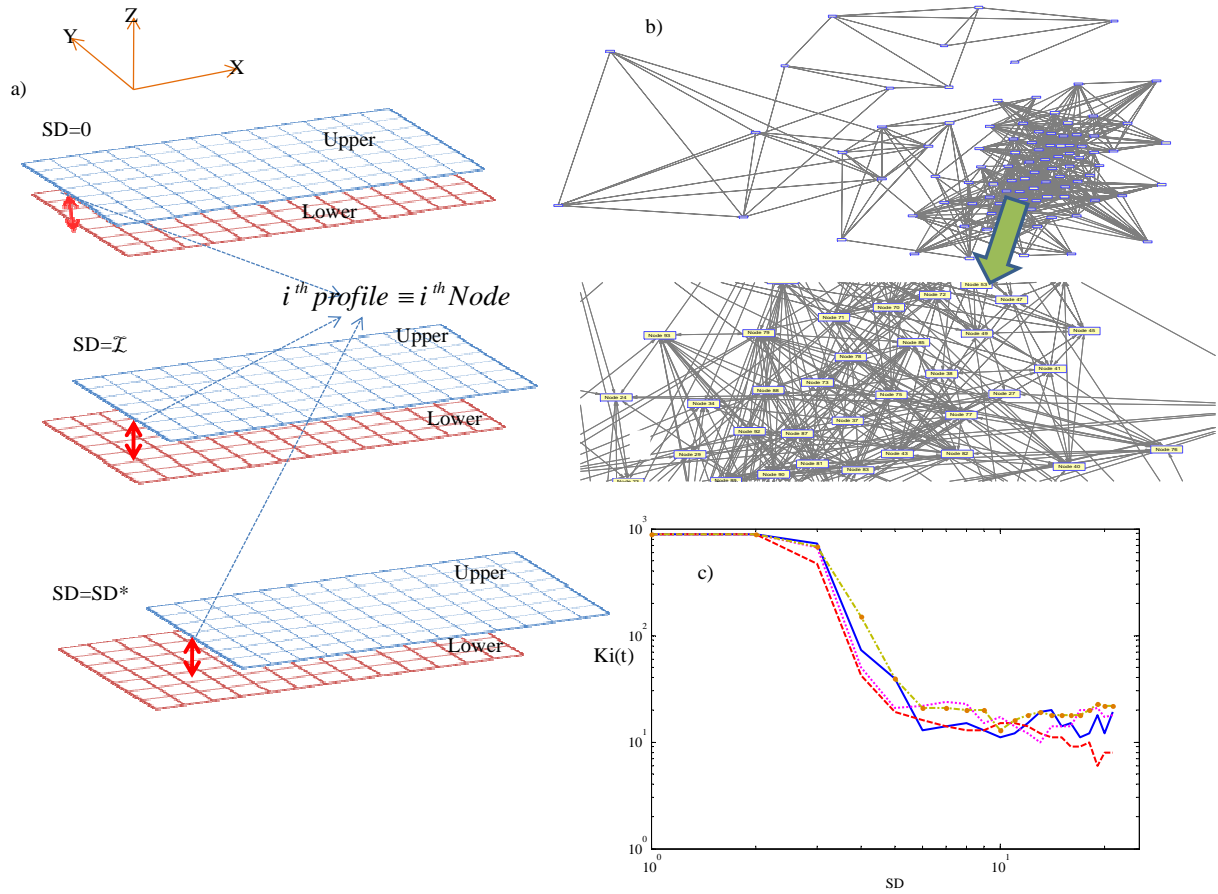

**Figure 1. a) Construction of a decaying complex network based on the measured apertures' profiles (the nodes references are coincided on the upper surface) ,b) Part of the created network (only first 100 nodes from 801 nodes) at 20 mm shear displacement (SD) and c) Evolutions of four nodes over the successive shear displacement (refer to the results section.)**

**Analysis of Aperture Evolution in a Rock Joint Using a Complex Network Approach-Figures**

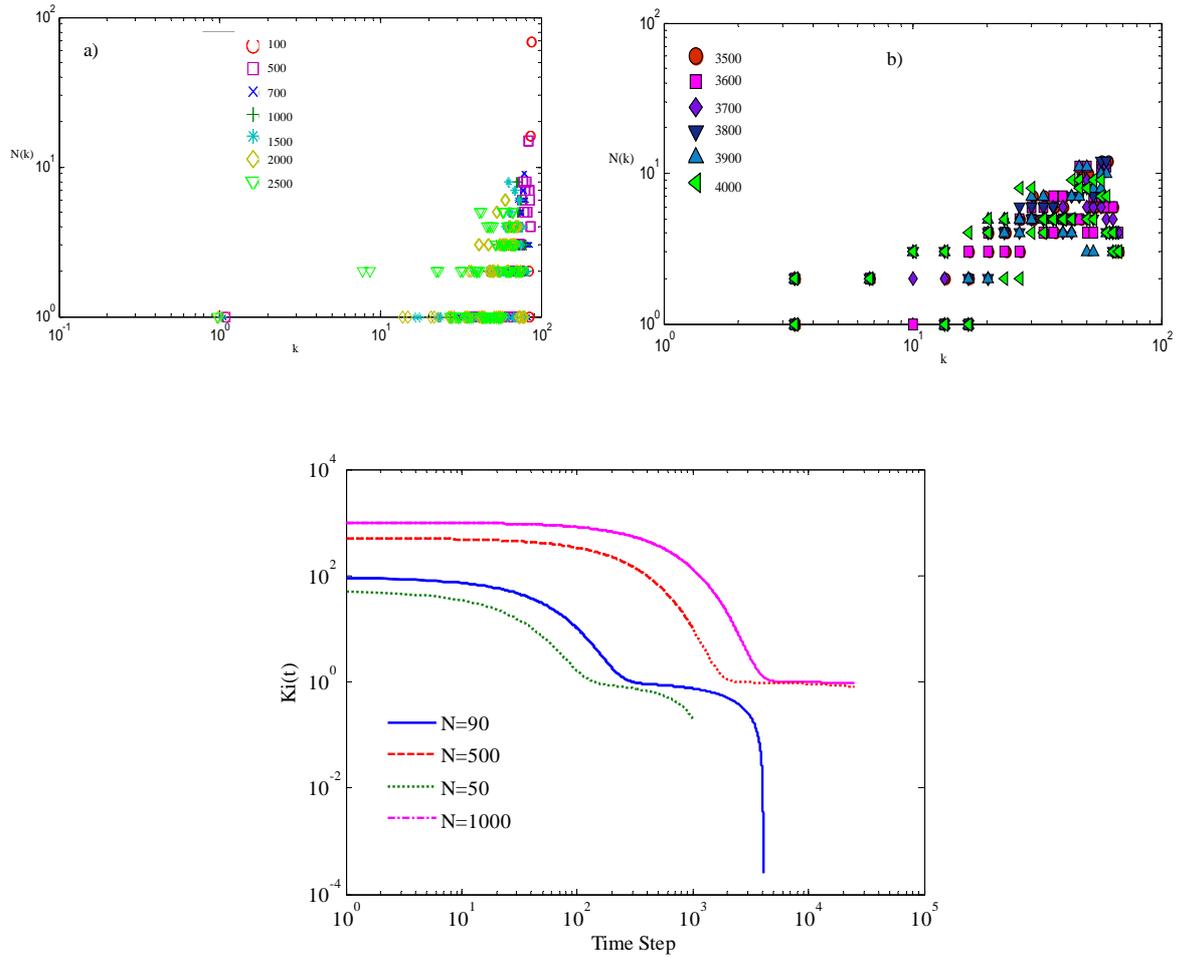

**Figure 2.** a&b) the evolution of the edges distributions using the proposed algorithm and on *N*=90; c) Analytical solution on several *N*

**Analysis of Aperture Evolution in a Rock Joint Using a Complex Network Approach-Figures**

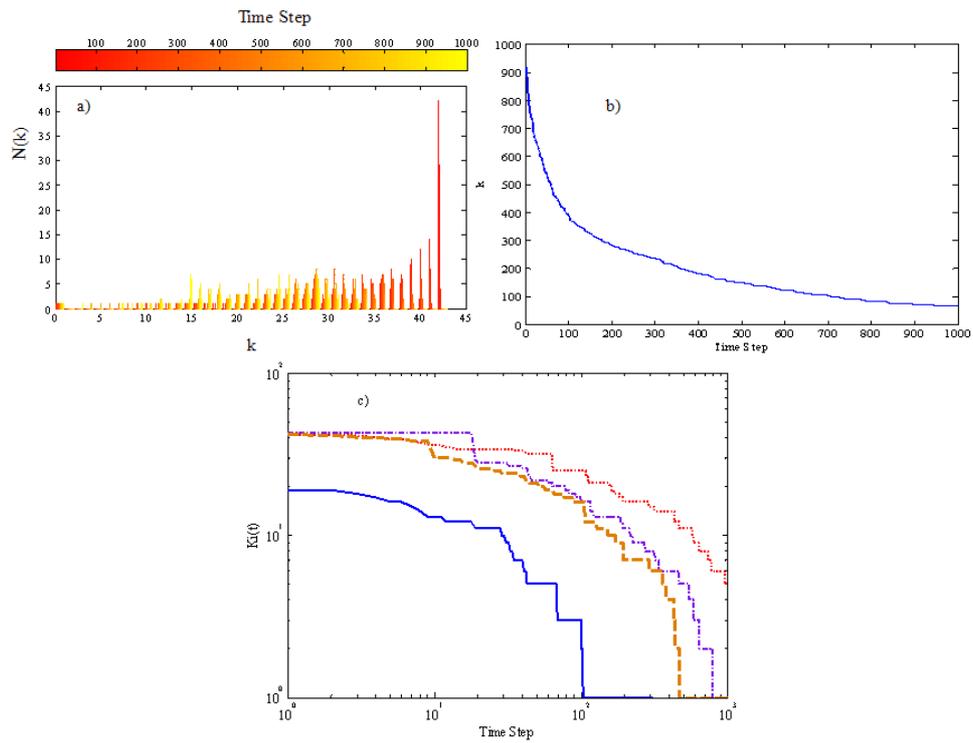

Figure 3. The evolution of the edges distributions and edge dependence using the proposed algorithm (a &b and c) on *N*=45

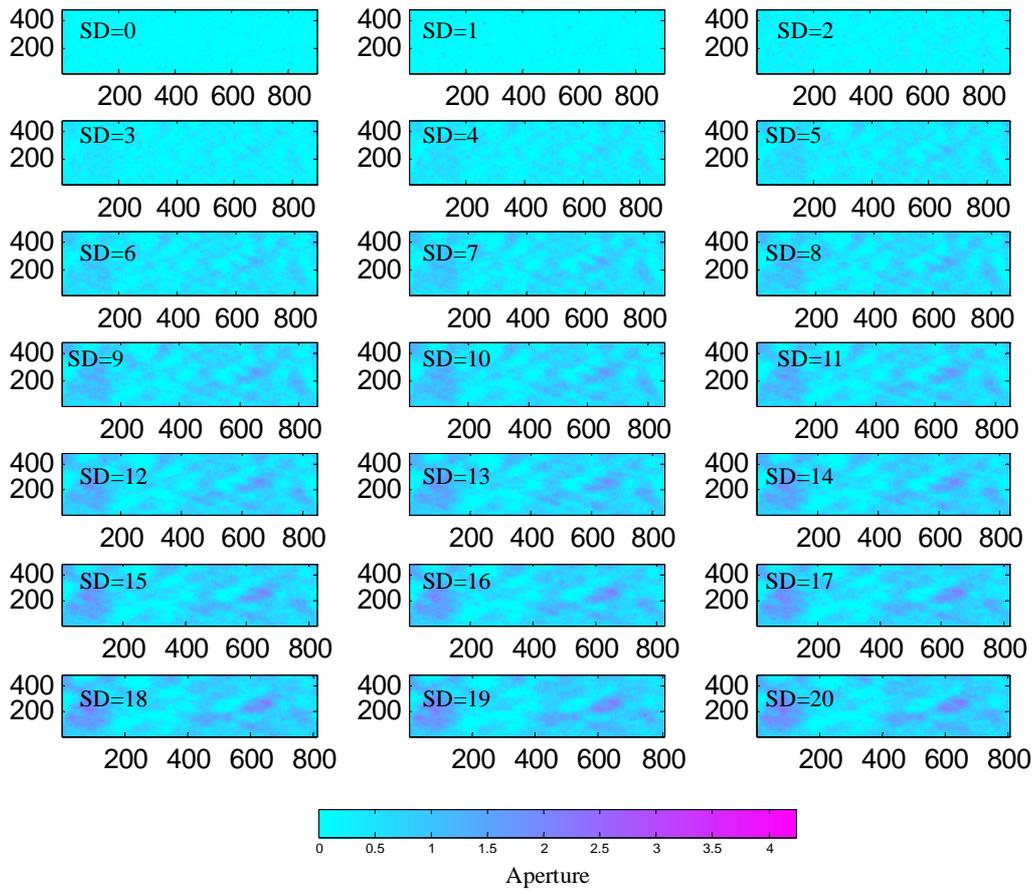

**Figure 4. The evolution of apertures 'patterns under successive shear displacements and 3 Mpa normal stresses (the axis show number of elements with a square element size of 0.2 mm)**

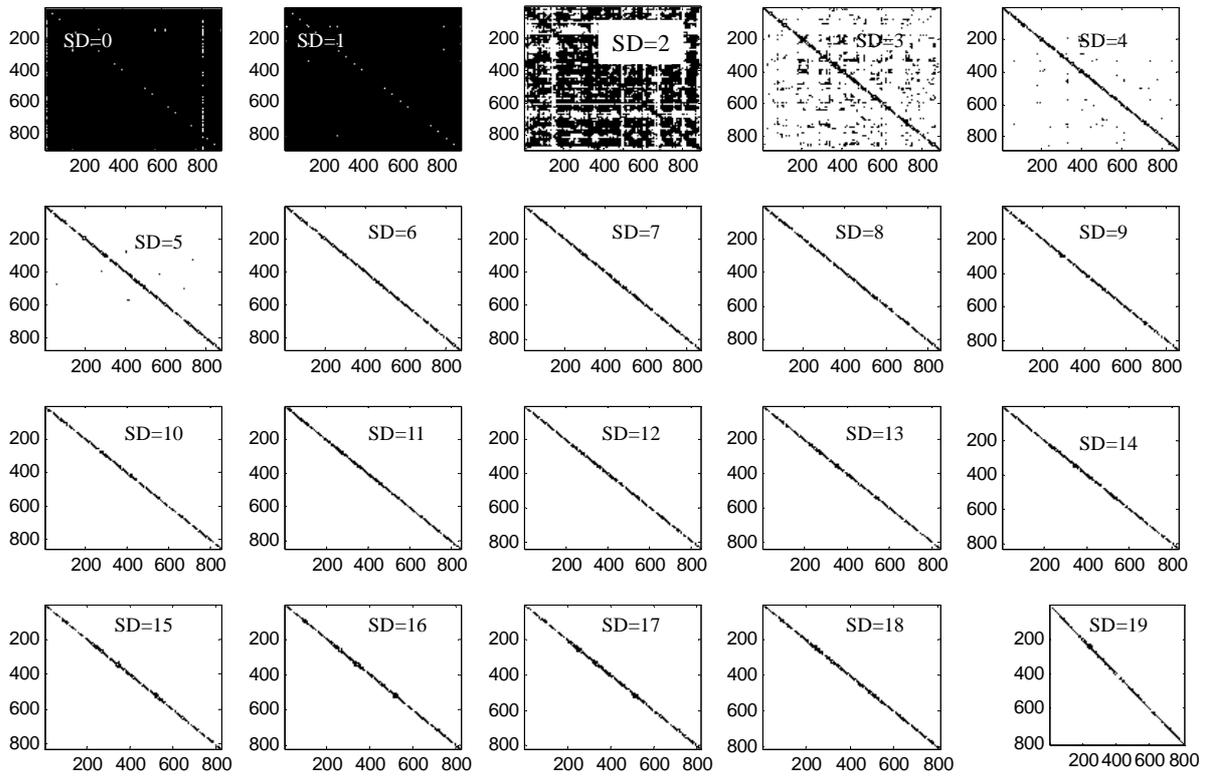

**Figure 5. The evolution of X-profiles networks (adjacency matrix visualization) using Euclidean distance and** $d \leq 5$

**Analysis of Aperture Evolution in a Rock Joint Using a Complex Network Approach-Figures**

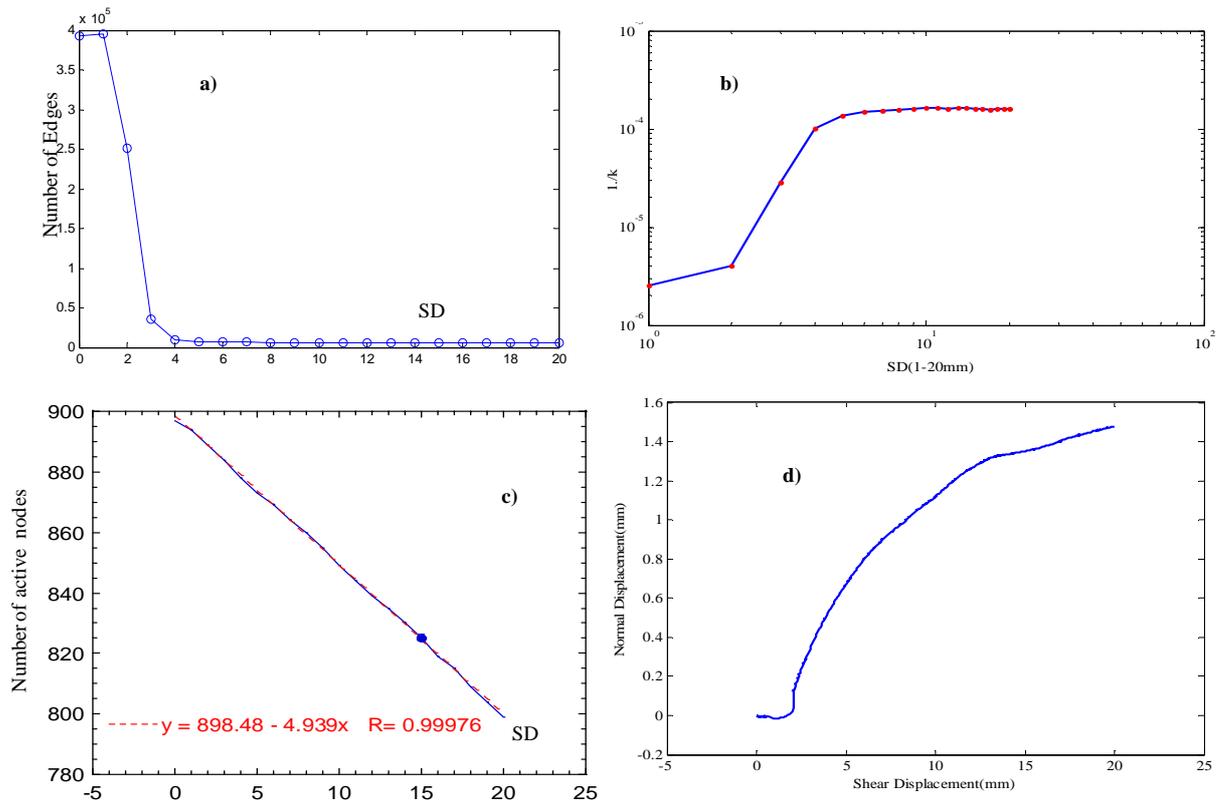

**Figure 6. Results on the X-profiles networks: a) Number of Edges -Shear displacement, b) Log-Log diagram of 1/Number Edges-Shear displacement, c) Number of active nodes (non-isolated vertices) -Shear displacement and d) Joint normal displacement-Shear displacement**

**Analysis of Aperture Evolution in a Rock Joint Using a Complex Network Approach-Figures**

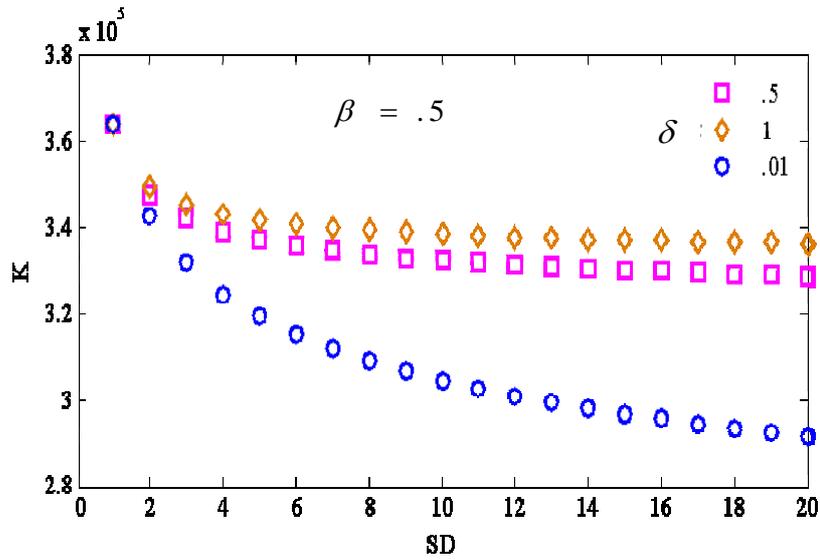

**Figure 7. Approximation of Number of Edges (over the X-profiles networks) using a modified sigmoid function** $k = 2 \times 10^5 \times (1 + e^{-\beta(1 + \frac{LnSD}{1 + \delta LnSD})})$

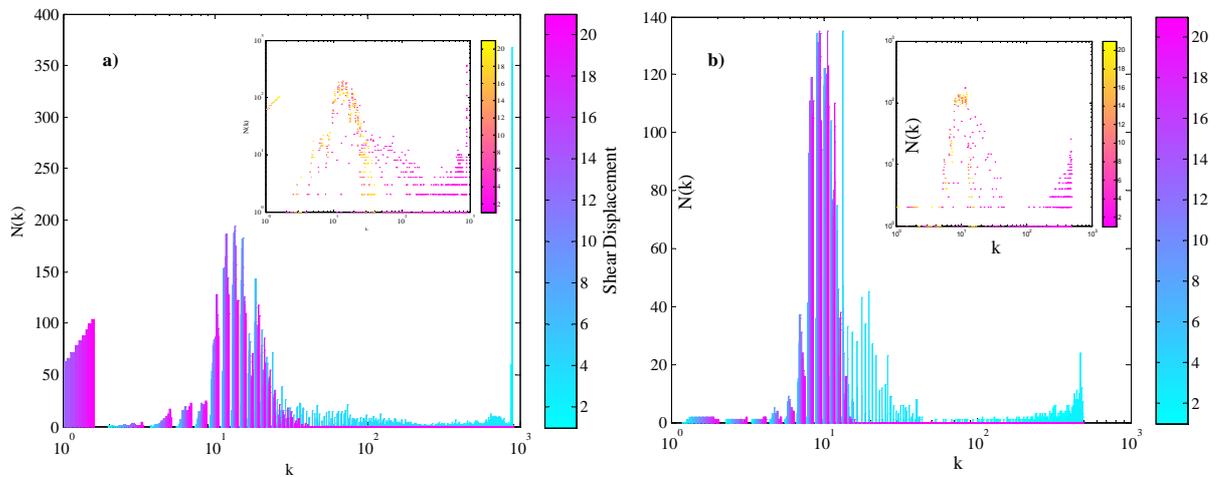

**Figure 8. Frequency of nodes connectivity evolution over the shear displacements on: a) X-profiles and b) Y-profiles; Inset: results in log-log coordinate**

**Analysis of Aperture Evolution in a Rock Joint Using a Complex Network Approach-Figures**

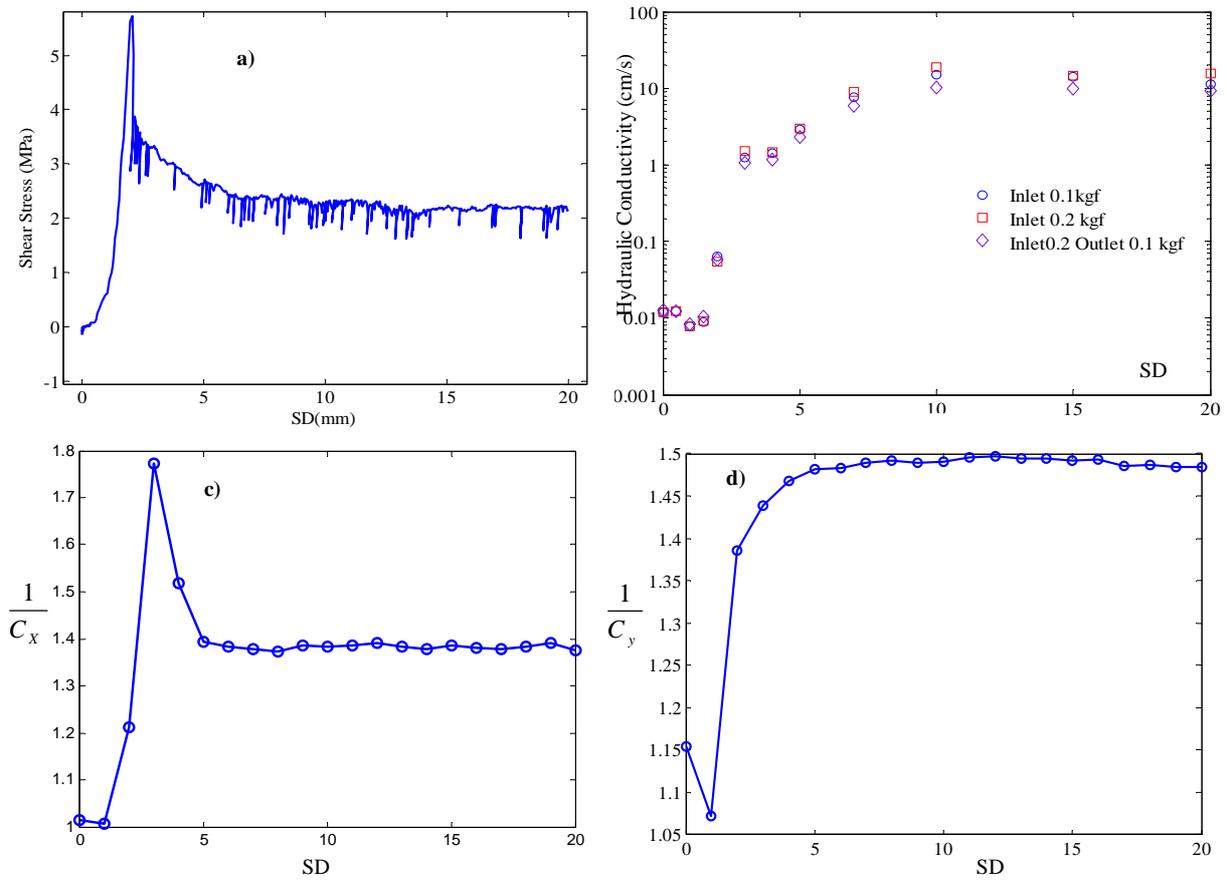

**Figure 9.a) Shear Stress -Shear Displacement (Normal Stress: 3Mpa), b) Hydraulic Conductivity - Shear Displacement associated with 3 different cases, c) &d) Inverse of Clustering Coefficients - Shear Displacement on the X and Y profiles ,respectively**

**Analysis of Aperture Evolution in a Rock Joint Using a Complex Network Approach-Figures**

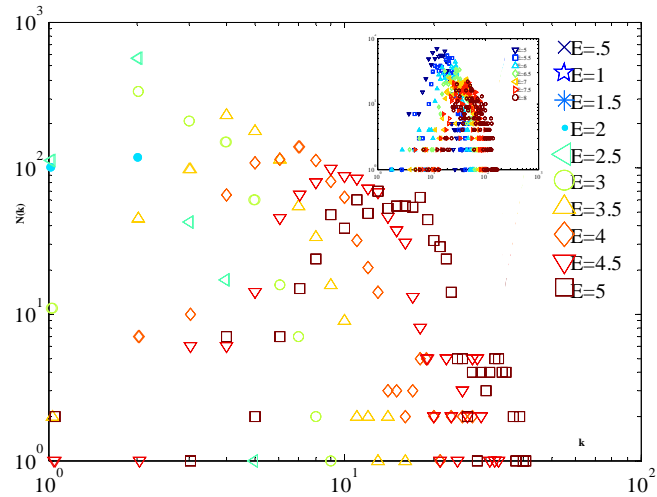

**Figure 10.** Effect of threshold variation ($0.5 \leq \xi \leq 5$) at SD=20mm –Euclidean distance (inset: $5 \leq \xi \leq 8$)

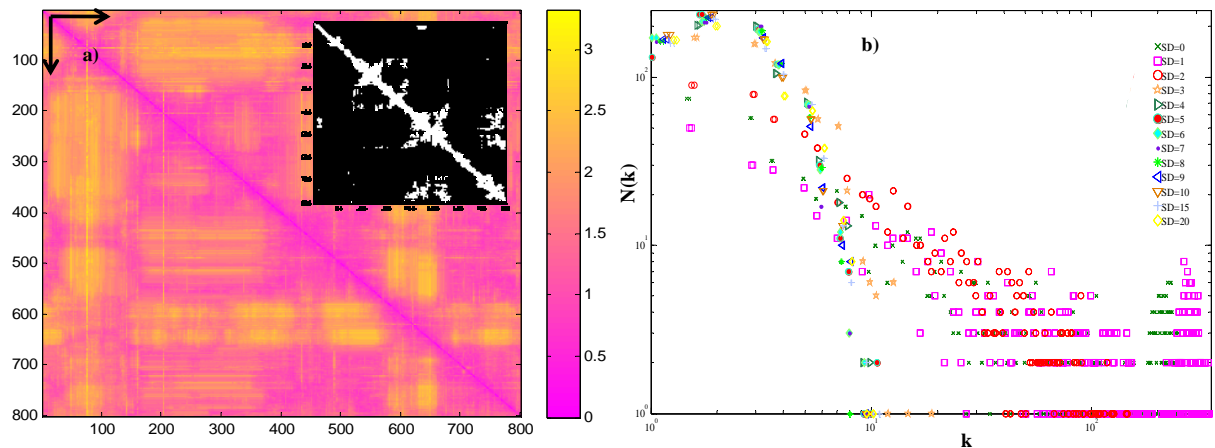

**Figure 11.** Results based on the Chebyshev distance: a) Distance matrix on X-profiles at 20 mm Shear Displacement (inset: Adjacency matrix of the proper network using $d \leq 1$) and b) transition of edges distribution to a Gaussian distribution ($d \leq .5$)

**Analysis of Aperture Evolution in a Rock Joint Using a Complex Network Approach-Figures**

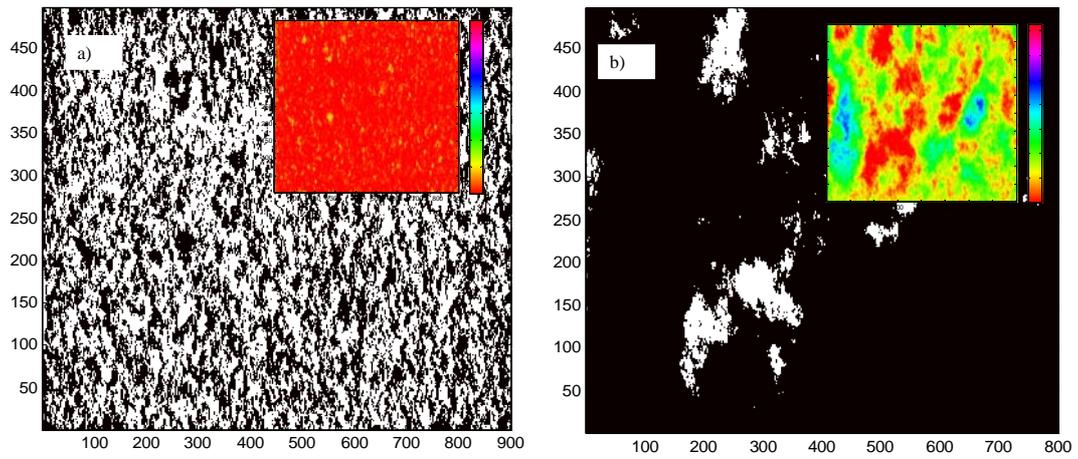

**Figure12. Discrimination of Contact zones (white hues) at a) SD=0 and b) SD=20-Insents are overall apertures patterns.**

**Analysis of Aperture Evolution in a Rock Joint Using a Complex Network Approach-Figures**

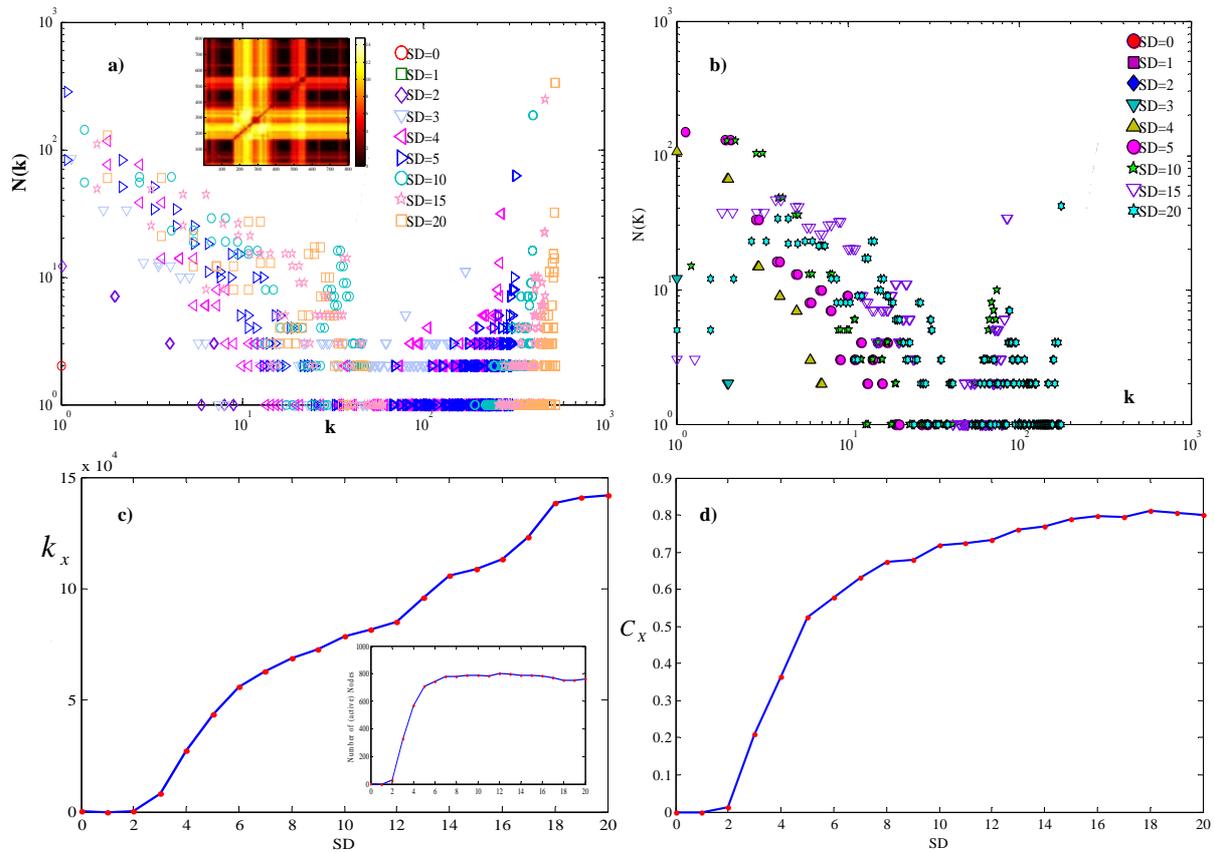

**Figure13.** Measurements on the contact zones (profile-profile) networks: a &b) Frequency of nodes connectivity evolution over the shear displacements on X-profiles (inset: appropriate distance matrix at 20 mm-SD) and Y-profiles, respectively, c) Growing of number of edges at X-profiles networks (inset: Number of active nodes evolving) and d) Clustering coefficient -Shear Displacement

11